\documentclass{nature}
\usepackage{graphicx}

\title{Investigating the nature of mass distribution surrounding the Galactic supermassive black hole}
 
\author{Man Ho Chan, Chak Man Lee, Chi Wai Yu}

\begin{document}
 
\maketitle
 
\begin{affiliations}
\item Department of Science and Environmental Studies, The Education University of Hong Kong, Tai Po, New Territories, Hong Kong, China
\end{affiliations}

\begin{abstract}
In the past three decades, many stars orbiting about the supermassive black hole (SMBH) at the Galactic Centre (Sgr A*) were identified. Their orbital nature can give stringent constraints for the mass of the SMBH. In particular, the star S2 has completed at least one period since our first detection of its position, which can provide rich information to examine the properties of the SMBH, and the astrophysical environment surrounding the SMBH. Here, we report an interesting phenomenon that if a significant amount of dark matter or stellar mass is distributed around the SMBH, the precession speed of the S2 stellar orbit could be `slow down' by at most 27\% compared with that without dark matter surrounding the SMBH, assuming the optimal dark matter scenario. We anticipate that future high quality observational data of the S2 stellar orbit or other stellar orbits can help reveal the actual mass distribution near the SMBH and the nature of dark matter.
\end{abstract}



\section{Introduction}
It is widely accepted that there exists a supermassive black hole (SMBH) at the centre of our Milky Way galaxy (Sgr A*). Based on the dynamics of the stars surrounding the SMBH, the mass of the SMBH is $M_{\rm BH} \approx (4.154 \pm 0.014) \times 10^6M_{\odot}$ \cite{Abuter}. Thank to the high quality observations, many stars surrounding the SMBH have been figured out \cite{Gillessen}. In particular, the S2 star has a very small period and it is almost the closest star to the SMBH. The S2 star is a young B-type main-sequence star with mass $M_{\rm S2} \approx 14-15M_{\odot}$ \cite{Habibi,Jovanovic}, which is one of the brightest members of the stars surrounding the SMBH \cite{Jovanovic}. Its motion has been monitored for almost 30 years by a few large telescopes in the world \cite{Eckart,Ghez98,Schodel,Eisenhauer,Ghez03,Ghez08,Heibel}. Based on current data, the orbital period of the S2 star is about 16 years \cite{Gillessen}. Its orbit has a very high eccentricity $e=0.88466 \pm 0.00018$ \cite{Abuter}. The pericentre and the apocentre of the orbit are 120 AU and 1820 AU respectively \cite{Abuter}. The S2 star has completed at least one period since our first close monitoring. Its orbital nature can give us important information in determining the mass of the SMBH and the astrophysical environment surrounding the SMBH \cite{Hees,Della}.

On the other hand, theories have predicted that dark matter distribution would be altered by the SMBH nearby. The collisionless behaviour of the dark matter particles would be likely to cause the adiabatic growth of the SMBH \cite{Gondolo}. Following the conservation of the angular momentum, dark matter particles would be accreted by the SMBH to form a dense spike or cusp-like structure \cite{Gondolo,Gnedin}. Such a high dark matter density spike has been widely examined based on radio and gamma-ray detections as a much larger rate of dark matter annihilation would be resulted \cite{Gondolo,Gnedin,Bertone,Fields,Shapiro,Chan}. 

Some recent studies have used the current S2 orbital data to constrain dark matter models \cite{Becerra,Heibel,Lacroix,Becerra2,Nampalliwar,Arguelles} or other alternative models of dark matter \cite{Jovanovic,Adkins}. However, a recent study has shown that the effect of the dark matter density spike is considerably weak for realistic dark matter densities \cite{Nampalliwar}. Therefore, it seems that no strong constraints of dark matter can be made based on the data of one complete revolution of the S2 star. Nevertheless, based on the analysis of the orbital precession of S2, we discover some important signatures that can reveal the properties of the dark matter density spike or the nature of the mass distribution around the SMBH. The orbital precession can be more precisely determined based on at least two complete orbital periods, which will be made in the coming two to three years.

Some previous studies have investigated the precession effect of the S2 orbit to estimate the mass of the SMBH \cite{Abuter}, and constrain the extended mass distribution \cite{Rubilar,Nucita,Heibel} or dark matter \cite{Arguelles,Zakharov,Dokuchaev,Dokuchaev2,Abuter2}. In this article, we extend the ideas of these studies and explicitly quantify the effect on the precession angle based on different dark matter or extended mass density profiles. We will further discuss how the astrophysical environment surrounding the SMBH or the nature of dark matter could be revealed based on the precession effect.

\section{Results}
Newtonian mechanics would give a perfect stable elliptical orbit for a star orbiting about the SMBH. However, as the mass of a SMBH is very large such that General Relativistic effect becomes important, a small non-linear term would appear in the equation of motion so that the resulting orbit would undergo the so-called Schwarzschild precession. For the SMBH at the Milky Way centre, the precession angle of the nearby stellar orbit depends on the SMBH mass $M_{\rm BH}$ and the stellar angular momentum $L$. The Schwarzschild precession angle for the S2 orbit is about $0.2^{\circ}$ per period (about $1.3^{\circ}$ per century). In the followings, we mainly follow the orbital parameters obtained by previous studies to perform the analysis. Generally, the uncertainties of the parameters used are very small so that the precession angle predicted is quite accurate. The data of the S2 orbit can be found in Table 1.

Now we assume that dark matter would distribute around the SMBH (the `SMBH+DM model') and the total enclosed mass inside the stellar orbit is $M(r)=M_{\rm BH}+M_{\rm DM}(r)$, where $M_{\rm DM}(r)$ is the enclosed dark matter mass and $r$ is the distance from the SMBH. Based on the S2 orbital data, the most optimistic estimated enclosed dark matter mass (or any extended mass) within the S2 orbit is smaller than 0.1\% of the SMBH mass \cite{Heibel,Abuter}. Therefore, we can use the method of perturbation to calculate the effect of the dark matter distribution surrounding the SMBH. We simply replace the point-mass term by the total enclosed mass $M(r)$ in the equation of motion. Here, since the enclosed dark matter mass depends on $r$, the non-linear effect would be significant after nearly one period of stellar motion. Such an effect would be manifested by the angle of precession. Note that the potential effect considered here is not limited to dark matter only. Any extended mass distribution (e.g. white dwarfs or neutron stars) could have the similar effect. However, we will particularly consider the case of dark matter as our major concern because theories predict that a concentrated dark matter density spike might be surrounding the SMBH \cite{Gondolo,Gnedin}, which might give significant impact on the precession angles of the surrounding stellar orbits.

Moreover, to compare the data with our analysis, we need to first transform the apparent orbit of the projected plane (in the line-of-sight direction) to the real orbital plane via three angles: the inclination angle, argument of pericenter and the ascending node angle \cite{Becerra2}. These angles can be constrained by current observational data \cite{Becerra2}. After transformation, we can obtain the best-fit orbit and perform analysis on the real orbital plane. Then we can transform all the results back to the apparent projected plane for illustrations. We take the data of the apparent orbit from a recent study of the S2 star \cite{Do}. We report the features of the precession analysis as follows.

\subsection{Dark matter versus no dark matter}
The precession could be influenced by the dark matter density spike significantly. In the `SMBH model', we assume that the S2 star orbiting about the SMBH only without dark matter. For the `SMBH+DM model', we assume the S2 star orbiting about the SMBH surrounding by the dark matter density spike distribution $\rho_{\rm DM} \propto r^{-\gamma}$. Theoretically, the cusp index $\gamma$ can range from 0.5 to 2.5 \cite{Fields}. Here, we consider one popular model - the Bahcall-Wolf cusp model \cite{Bahcall} - to describe the dark matter density spike distribution for illustration. The cusp-like structure of the density profile in the model is consistent with the predicted dark matter density spike distribution \cite{Gondolo,Gnedin}. The cusp index $\gamma=7/4$ is close to the average value of the theoretical possible range, although the Bahcall-Wolf cusp model is not proposed to describe the dark matter density profile originally. The optimal parameters of the model can be constrained by the current S2 data of one period \cite{Heibel,Abuter}. We model the precession angles after 50 periods for both of the `SMBH model' and `SMBH+DM model'. By averaging after 50 periods, the precession angles per one period on the real orbital plane are $0.2077^{\circ}$ and $0.1515^{\circ}$ for the `SMBH model' and `SMBH+DM model' respectively (see Fig.~1 for the orbits on the apparent projected orbital plane). Therefore, the dark matter distribution seems like `slowing down' the precession speed of the S2 orbit by 27\%. If such an angle is observed, this could be regarded as an indirect evidence of the existence of dark matter (or similar extended mass distribution) surrounding the SMBH. Some studies also suggest $\gamma=1.5$ due to the interaction between dark matter and baryons \cite{Gnedin}. By assuming the same dark matter content, the precession angle per one period for this profile is 0.1547, which is slightly larger than the one following the Bahcall-Wolf model. Generally, a larger value of $\gamma$ would give a smaller precession angle. On the other hand, this method can also be applied in testing General Relativity \cite{Hees,Do,Iorio} or modified gravity models \cite{Jovanovic,Adkins}. Note that the mass distribution due to neutron stars or white dwarfs can also cause similar effect on precession. Recent studies show that the fiducial model of neutron star distribution at the Galactic Centre gives the central mass density $<2\times 10^{-12}$ kg m$^{-3}$ \cite{Generozov}, which is ten times smaller than the central mass density assumed in the Bahcall-Wolf cusp model. Therefore, we simply assume that dark matter dominates the mass distribution near the SMBH. Nevertheless, it is still possible that the mass distribution constrained by this method originates from baryonic matter, but not dark matter.

\subsection{Constraining the mass distribution}
Since the precession angle can be influenced by dark matter or baryonic matter, the actual precession angle observed could be used to differentiate different models of dark matter density distribution or baryonic matter distribution. In fact, the functional form and the parameters involved (e.g. index of the density spike) for the mass distribution surrounding the SMBH are uncertain. There are some variations in the functional form of the mass density distribution based on the theoretical predictions \cite{Fields,Nucita}. Here, we model the precession angles of the S2 orbit based on two popular density models with two entirely different functional forms, the `Plummer model' \cite{Plummer} and the `Bahcall-Wolf cusp model' \cite{Bahcall}. The Plummer model has a constant density core at the centre and it is commonly modeled as the distribution of a stellar cluster. The Bahcall-Wolf cusp model has a central density cusp and it is close to the prediction of the dark matter density spike distribution. By considering the optimal scenarios for both models \cite{Heibel,Abuter}, we find that the resultant precession angles are slightly different for different models. The precession angle on the real orbital plane for the Plummer model is $0.1654^{\circ}$, which is slightly larger than that for the Bahcall-Wolf cusp model (see Fig.~1 and Table 2). Since the cusp property in the Bahcall-Wolf cusp model gives a higher dark matter density in the inner central region, this implies that a higher central density would give a larger `slowing down effect' on the precession angle of the S2 orbit. Moreover, if the value of the precession angle can be determined accurately, it could give some constraints on the density distribution model or the cusp index $\gamma$. This can also help us review our understanding of the dynamics of particle dark matter, which has not been rigorously tested.

\subsection{The annihilation effect}
Some particle physics models predict that dark matter particles can self-annihilate to give high-energy particles \cite{Profumo}. The dark matter annihilation rate is directly proportional to the square of dark matter density. Since the inner dark matter density would be very high due to the cusp distribution, the central annihilation rate would also be very high. As a result, a large amount of dark matter particles would be annihilated in the inner region so that the central dark matter density would be subsequently much smaller. The final dark matter density would achieve a so-called `annihilation plateau' in the innermost region of the density spike when the annihilation rate is high enough \cite{Fields}. This would suppress the dark matter effect on the precession angle. For example, assuming the mass of a dark matter particle $m_{\rm DM}=1$ TeV and following the standard thermal annihilation cross section $\langle \sigma v \rangle=2.2 \times 10^{-26}$ cm$^3$/s, the precession angle per one period on the real orbital plane for the Bahcall-Wolf cusp model would increase to $0.1774^{\circ}$. Generally speaking, when the annihilation rate per unit dark matter mass is large, the influence of dark matter becomes less important and the precession angle would approach the precession angle of the `SMBH model' ($0.2077^{\circ}$). We plot the variation of the precession angle against $m_{\rm DM}/ \langle \sigma v \rangle$ in Fig.~2. Therefore, the precession angle of the S2 orbit could provide some hints on the annihilation parameters. The results can also be combined with the radio analysis \cite{Bertone} and gamma-ray analysis \cite{Fields,Shapiro} to come up with a more stringent constraint on the annihilation parameters. 

\section{Discussion}
In this study, we have explicitly quantified the effect of dark matter or any extended mass distribution on the precession angle of the S2 orbit theoretically. We have shown that dark matter distributing with the Bahcall-Wolf cusp model surrounding the SMBH can give a significant smaller precession angle for the S2 orbit. In the optimal dark matter scenarios, the precession speed can be smaller by at most 27\% compared with that without dark matter surrounding the SMBH. Some previous studies have considered the precession angle of the S2 orbit to constrain the mass of the SMBH \cite{Abuter,Abuter2}, and the extended mass distribution (including dark matter) surrounding the SMBH \cite{Rubilar,Nucita,Heibel,Arguelles,Zakharov,Dokuchaev,Dokuchaev2,Abuter2}. Nevertheless, we have shown in this study that the precession angles for different extended mass distributions would be significantly different. Since the stellar density distribution is close to the Plummer model's profile while the dark matter density distribution is close to the Bahcall-Wolf cusp model's profile, observing the actual precession angle of the S2 star can help differentiate the nature of the extended mass distribution (stellar vs. dark matter). Based on our analysis, such an effect on the precession angle could be significant and easily noticed from future observational data (after at least two complete orbital periods). Moreover, dark matter annihilation would also affect the value of the precession angle, which has not been realised and discussed before. Since dark matter annihilation would wash out the cusp properties of the dark matter density spike, this effect could be manifested by the change in the precession angle. Therefore, the annihilation parameters (the annihilation cross section per unit dark matter mass) can be theoretically constrained by this method. Generally speaking, non-annihilating dark matter with cusp-like density distribution nearby the SMBH would give the largest effect on the precession angle (i.e. the smallest precession angle per period).

Nevertheless, the constraints obtained from the precession angle are model-dependent. The density profiles assumed in this study are just examples for investigation. The actual cusp index or the actual functional form might be different from what we have assumed. In view of this, the single value of the precession angle of the S2 orbit might not be able to differentiate the effects from different possible models. More data would be needed for giving differentiation among different models. For example, there are some newly discovered stars (S62, S4711, S4714 and S4716) inside the S2 orbit which have orbital periods less than 12 years \cite{Peibker2,Peibker3,Peibker}. The dark matter or extended mass distribution would also affect these stars so that the data of their orbits can also be used to examine the mass distribution surrounding the SMBH in the coming decade. Therefore, combining these information would give us a clearer astrophysical picture about the environment of the SMBH and help reveal the dark matter properties.

\begin{table}
\caption{Parameters of the S2 orbit}

\begin{tabular}{ |l|l|}
 \hline\hline
  Mass of S2 star& 14-15$M_\odot$\cite{Habibi,Jovanovic} \\
  $M_{\rm BH}$& $(4.154\pm0.014)\times 10^6M_\odot$ \cite{Abuter} \\
  Orbit period &15.9-16.0 yrs \cite{Hees} \\
  Eccentricity ($e$)  &0.88466 $\pm$ 0.00018 \cite{Abuter} \\
  Schwarzschild radius ($r_s=2GM_{\rm BH}/c^2$)  & 0.08 AU \cite{Abuter1,Abuter3}\\
  Pericentre  & 120 AU \cite{Abuter}\\
  Apocentre  & 1820 AU \cite{Abuter} \\
  Velocity at pericentre & 7650 km/s \cite{Abuter1}\\
  Inclination $i$ & $134.3533^{\circ}$ \cite{Becerra2}\\
  Argument of pericentre $\omega$ & $66.7724^{\circ}$ \cite{Becerra2}\\
  Ascending node $\Omega$ & $228.024^{\circ}$ \cite{Becerra2}\\
 \hline\hline
\end{tabular}
\end{table}

\begin{table}
\caption{Fitted parameters of the S2 orbit for the `SMBH model' and the `SMBH+DM' model.}
\begin{tabular}{ |l|l|l|l|l|}
 \hline\hline
Parameter&  BH only &Plummer model& Cusp model & Cusp model \\
         &          &             & ($\gamma=1.5$)            & ($\gamma=7/4$) \\
\hline
Scale density $\rho_0$ (kg/m$^3$) &  - & $1.69 \times 10^{-10}$ \cite{Heibel} & $2.88\times 10^{-11}$ & $2.24\times 10^{-11}$ \cite{Heibel} \\
Scale radius $r_0$ (pc) & - & $0.012$ \cite{Abuter} & $0.012$ \cite{Abuter} & $0.012$ \cite{Abuter} \\
Precession angle per one period (deg) & 0.2077 &  0.1654  & 0.1547 & 0.1515\\
  \hline\hline
\end{tabular}
\end{table}

\begin{figure}
\vskip 3mm
\includegraphics[width=100mm]{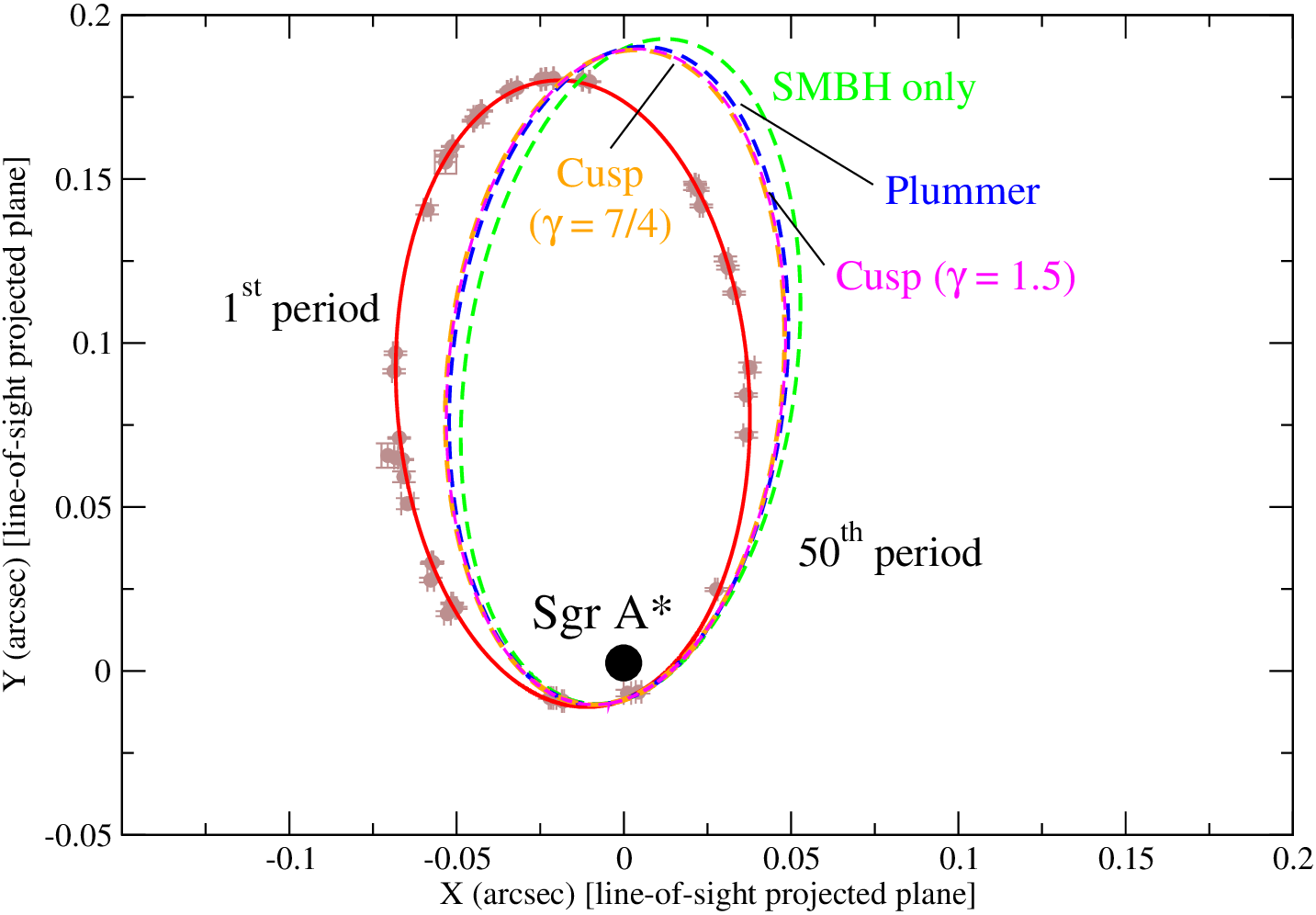}
\caption{The red ellipse represents the best-fit orbit of the S2 star based on the observational data in brown dots with error bars \cite{Do} on the apparent projected orbital plane (XY-plane). The green, orange, magenta, and blue dashed lines respectively represent the predicted orbits of the S2 star after the 50$^{\rm th}$ period for the `SMBH model', `SMBH+DM model' with the Bahcall-Wolf cusp distribution ($\gamma=7/4$), `SMBM+DM model' with the cusp distribution $\gamma=1.5$, and the `SMBH+DM model' with the Plummer distribution (transformed to the apparent projected orbital plane). The position of the SMBH (Sgr A*) is at $(X,Y)=(-0.000083",0.0024893")$ \cite{Becerra}.}
\label{Fig1}
\vskip 3mm
\end{figure}

\begin{figure}
\vskip 3mm
\includegraphics[width=100mm]{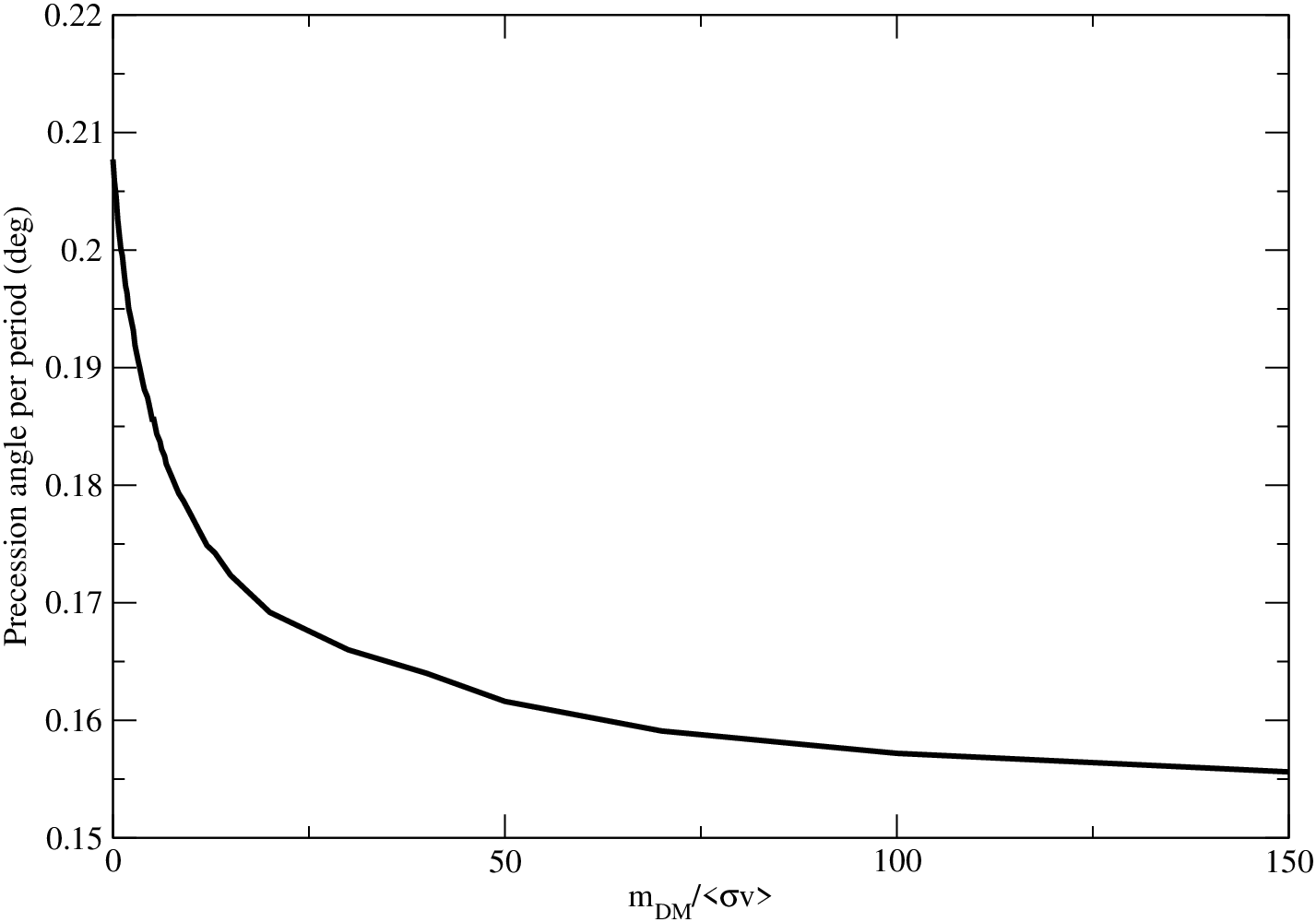}
\caption{The black solid line indicates the variation of the precession angle of the S2 star per period against the parameter $m_{\rm DM}/ \langle \sigma v \rangle$. Here, the value of the $m_{\rm DM}/ \langle \sigma v \rangle$ is normalised by the ratio $100~{\rm GeV}/2.2 \times 10^{-26}~{\rm cm^3/s}$.}
\label{Fig2}
\vskip 3mm
\end{figure}

\begin{methods}
\subsection{The equation of motion around the SMBH}
For Schwarzschild black hole model, the spherical symmetric space-time metric can be written as \cite{Arguelles}
\begin{equation}
ds^2=A(r)c^2dt^2-B(r)dr^2-r^2(d\theta^2+\sin^2\theta d\phi^2)
\label{metric}
\end{equation}
where $(r, \theta,\phi)$ are the spherical coordinates, $A(r)=1-r_s/r$ with $r_s=2GM_{\rm BH}/c^2$, and $B(r)=1/A(r)$. The equation of motion of a star orbiting about a SMBH in the space-time metric, assuming without loss of generality the motion on the fixed plane $\theta=\pi/2$, is given by:
\begin{equation}
\frac{d^2u}{d\phi^2}+u=\frac{GM_{\rm BH}}{L^2}+3\frac{GM_{\rm BH}}{c^2}u^2,
\label{trajectory}
\end{equation}
where $u=1/r$, $L=rv_{\phi}$ is the angular momentum, and $v_{\phi}$ is the velocity at the pericentre. By using the perturbation method, the solution of $u(\phi)$ can be approximately given by 
\begin{equation}
u(\phi) \approx \frac{GM_{\rm BH}}{L^2}[1+e\cos(\phi-\epsilon\phi)],
\end{equation}
where $e$ and $\epsilon$ are constant. Therefore, the precession angle of the stellar orbit moving about a pure SMBH is approximately 
\begin{equation}
\Delta\phi_{\rm precession}=2\pi\epsilon=6\pi \frac{G^2M_{\rm BH}^2}{c^2L^2}.
\label{precession}
\end{equation}
Using the data of the S2 star (see Table 1), the Schwarzschild precession angle is about $0.2^{\circ}$.

In the presence of dark matter surrounding the SMBH, following the method of perturbation, the SMBH mass $M_{\rm BH}$ in Eq. (\ref{trajectory}) could be replaced by
\begin{equation}
M(r)=M_{\rm BH}+M_{\rm DM}(r),
\end{equation}
where $M_{\rm DM}(r)$ is the enclosed dark matter mass. This can be done because $M_{\rm DM}(r)$ is much smaller than $M_{\rm BH}$ for the most optimistic dark matter distribution constrained by the S2 orbital data \cite{Heibel}. Therefore, the equation of motion finally becomes 
\begin{equation}
\frac{d^2u}{d\phi^2}+u=\frac{GM(u)}{L^2}+3\frac{GM(u)}{c^2}u^2.
\label{trajectory1}
\end{equation}
The final orbit $r(\phi)=1/u(\phi)$ of the S2 star could be obtained by solving Eq.~(\ref{trajectory1}) numerically.

\subsection{Dark matter density model}
The dark matter density spike surrounding a SMBH is commonly modelled by a cusp model:
\begin{equation}
\rho_{\rm DM}(r)=\rho_0 \left(\frac{r_0}{r} \right)^{\gamma},
\end{equation}
where $\rho_0$, $r_0$ and $\gamma$ are the scale density parameter, scale radius parameter and the cusp index respectively. The cusp index is model-dependent while the parameters $\rho_0$ and $r_0$ can be fitted empirically by the data of the S2 orbit. For the Bahcall-Wolf cusp model \cite{Bahcall} considered in our analysis, the cusp index is $\gamma=7/4$. The optimal values of parameters fitted by the data are $\rho_0=2.24 \times 10^{-11}$ kg/m$^3$ \cite{Heibel} and $r_0=0.012$ pc \cite{Abuter}. Moreover, we also test the cusp model with $\gamma=1.5$. For the same dark matter content inside the S2 orbit, by keeping the same scale radius $r_0=0.012$ pc, the optimal scale density parameter is $\rho_0=2.88\times 10^{-11}$ kg/m$^3$.

Beside the cusp model, the Plummer model \cite{Plummer} is another benchmark density model usually assumed at the Galactic Centre \cite{Nucita}. The density for the Plummer model is
\begin{equation}
\rho_{\rm DM}(r)=\rho_0 \left(1+\frac{r^2}{r_0^2} \right)^{-5/2}.
\end{equation}
The Plummer model is commonly modeled as the density distribution of a stellar cluster. Here, it can also be viewed as a cored dark matter density profile or any baryonic matter distribution for comparison. The optimal values fitted by the data of the S2 orbit are $\rho_0=1.69 \times 10^{-10}$ kg/m$^3$ \cite{Heibel} and $r_0=0.012$ pc \cite{Abuter}. The enclosed dark matter or extended mass for different dark matter or any extended mass distribution is thus given by
\begin{equation}
M_{\rm DM}(r)=\int_{4GM_{\rm BH}/c^2}^r  \rho_{\rm DM}(r')4\pi r'^2dr'.
\label{MDM}
\end{equation}

If dark matter would self-annihilate, the annihilation rate would be proportional to the square of dark matter density. Therefore, the dark matter density in the innermost region would decrease significantly. The original dark matter density distribution $\rho_{\rm DM}(r)$ would be modified by the following dark matter annihilation plateau density distribution:
\begin{equation}
\rho_{\rm DM,ann}(r)=\frac{\rho_{\rm DM}(r)\rho_{\rm in}(t,r)}{\rho_{\rm DM}(r)+\rho_{\rm in}(t,r)},
\end{equation}
where
\begin{equation}
\rho_{\rm in}(t,r)=\left(\frac{m_{\rm DM}}{\langle \sigma v \rangle t} \right)\left(\frac{r}{r_{\rm in}} \right)^{-0.5},
\end{equation}
with $t \approx 10^{10}$ yrs is the age of the SMBH, $\langle \sigma v \rangle$ is the annihilation cross section, $m_{\rm DM}$ is the mass of a dark matter particle, and $r_{\rm in} \approx 3.1 \times 10^{-3}$ pc \cite{Gnedin,Fields}. In standard cosmology, the thermal annihilation cross section is $\langle \sigma v \rangle=2.2 \times 10^{-26}$ cm$^3$/s for $m_{\rm DM} \ge 10$ GeV \cite{Steigman}. 

\subsection{Transformation of the apparent orbit}
The S2 orbit observed is on the apparent projected plane (XY-plane) along the line-of-sight direction. To get a better comparison of the orbit calculated from the equation of motion, we need to transform the apparent projected orbital plane to the real orbital plane (xy-plane) \cite{Heibel,Becerra2}. 

The Cartesian coordinate transformation from the XY-plane of the apparent orbit to the xy-plane of the real orbit can be done via the following relation \cite{Becerra2}:

\begin{equation}
\left(
\begin{array}{l}
X \\
Y   
\end{array}
\right)=
\left(
\begin{array}{ll}
A_{11} & A_{12} \\
A_{21} & A_{22} 
\end{array}
\right)
\left(
\begin{array}{l}
x \\
y 
\end{array}
\right)
\end{equation}

with
\begin{equation}
\left(
\begin{array}{ll}
A_{11} & A_{12} \\
A_{21} & A_{22}
\end{array}
\right)=
\left(
\begin{array}{ll}
\sin\Omega\cos\omega+\cos\Omega\sin\omega\cos i & -\sin\Omega\sin\omega+\cos\Omega\cos\omega\cos i \\
\cos\Omega\cos\omega-\sin\Omega\sin\omega\cos i & -\cos\Omega\sin\omega-\sin\Omega\cos\omega\cos i
\end{array}
\right)
\end{equation}
where $\omega$, $i$ and $\Omega$ are the osculating orbital elements, respectively representing the argument of pericentre, the inclination between the real orbit and the observation plane, and the ascending node angle \cite{Becerra2}. The osculating orbit elements constrained by the S2 data are shown in Table 1. The coordinates of the SMBH are transformed from $(X_0,Y_0)=(-0.000083",0.0024893")$ to $(x_0,y_0)=(0,0)$. 
\end{methods}
 


\begin{thebibliography}{1}
\bibitem{Abuter} Abuter R. {\it et al.} [GRAVITY Collaboration], Detection of the Schwarzschild precession in the orbit of the star S2 near the Galactic centre massive black hole, Astron. Astrophys. 636, L5 (2020).
\bibitem{Gillessen} Gillessen S. {\it et al.}, An update on monitoring stellar orbits in the Galactic Center, Astrophys. J. 837, 30 (2017).
\bibitem{Habibi} Habibi M. {\it et al.}, Twelve years of spectroscopic monitoring in the galactic center: The closest look at S-stars near the black hole, Astrophys. J. 847, 120 (2017).
\bibitem{Jovanovic} Jovanovi\'c P., Borka D., Borka Jovanovi\'c V., Zakharov A. F., Influence of bulk mass distribution on orbital precession of S2 star in Yukawa gravity, Eur. Phys. J. D 75, 145 (2021).
\bibitem{Adkins} Adkins G. S. \& McDonnell J., Orbital precession due to central-force perturbations, Phys. Rev. D 75, 082001 (2007).
\bibitem{Eckart} Eckart A., Genzel R., Observations of stellar proper motions near the Galactic Centre, Nature 383, 415 (1996).
\bibitem{Ghez98} Ghez A. M., Klein B. L., Morris M., Becklin E. E., High proper-motion stars in the vicinity of Sagittarius A*: Evidence for a supermassive black hole at the center of our Galaxy, Astrophys. J. 509, 678 (1998).
\bibitem{Schodel} Sch\"odel R., Ott T., Genzel R. {\it et al.}, A star in a 15.2-year orbit around the supermassive black hole at the centre of the Milky Way, Nature 419, 694 (2002).
\bibitem{Eisenhauer} Eisenhauer F., Sch$\ddot{\rm o}$del R., Genzel R., Ott T., Tecza, M., Abuter R., Eckart A., Alexander T., A geometric determination of the distance to the Galactic Center, Astrophys. J. 597, L121 (2003).
\bibitem{Ghez03} Ghez A. M., Duch\^ene G., Matthews K. {\it et al.}, The first measurement of spectral lines in a short-period star bound to the galaxy's central black hole: A paradox of youth, Astrophys. J. 586, L127 (2003).
\bibitem{Ghez08} Ghez A. M., Salim S., Weinberg N. N. {\it et al.}, Measuring distance and properties of the Milky Way's central supermassive black hole with stellar orbits, Astrophys. J. 689, 1044 (2008).
\bibitem{Heibel} Hei$\beta$el G., Paumard T., Perrin G., Vincent F., The dark mass signature in the orbit of S2, Astron. Astrophys. 660, A13 (2022).
\bibitem{Hees}Hees A. et al, Testing general relativity with stellar orbits around the supermassive black hole in our galactic center, Phys. Rev. Lett. 118, 211101 (2017).
\bibitem{Della} Della Monica R., de Martino I., Unveiling the nature of SgrA* with the geodesic motion of S-stars, J. Cosmol. Astropart. Phys. 03, 007 (2022).
\bibitem{Gondolo} Gondolo P., Silk J., Dark matter annihilation at the galactic center, Phys. Rev. Lett. 83, 1719 (1999).
\bibitem{Gnedin} Gnedin, O. Y., Primack, J. R., Dark matter profile in the Galactic Center, Phys. Rev. Lett. 93, 061302 (2004).
\bibitem{Bertone} Bertone G., Sigl G., Silk J., Annihilation radiation from a dark matter spike at the galactic centre, Mon. Not. R. Astron. Soc. 337, 98 (2002).
\bibitem{Fields} Fields B. D. , Shapiro S. L., Shelto J., Galactic center gamma-ray excess from dark matter annihilation: is there a black hole spike? Phys. Rev. Lett. 113, 151302 (2014).
\bibitem{Shapiro} Shapiro S. L., Shelton J., Weak annihilation cusp inside the dark matter spike about a black hole, Phys. Rev. D 93, 123510 (2016).
\bibitem{Chan} Chan M. H., Constraining the population of intermediate-mass black holes by gamma-ray data of the Fornax cluster, Mon. Not. R. Astron. Soc. 481, 3618 (2018).
\bibitem{Becerra} Becerra-Vergara E. A., Arguelles C. R., Krut A., Rueda J. A., Ruffini R., Hinting a dark matter nature of Sgr A* via the S-stars, Mon. Not. R. Astron. Soc. 505, L64 (2021).
\bibitem{Lacroix} Lacroix T., Dynamical constraints on a dark matter spike at the Galactic Centre from stellar orbits, Astron. Astrophys. 619, A46 (2018).
\bibitem{Becerra2} Becerra-Vergara E. A., Arguelles C. R., Krut A., Rueda J. A., Ruffini R., The geodesic motion of S2 and G2 as a test of the fermionic dark matter nature of our galactic core, Astron. Astrophys. 641, A34 (2020).
\bibitem{Nampalliwar} Nampalliwar S., Saurabh K., Jusufi K., Wu Q., Jamil M., Salucci P., Modelling the Sgr A* black hole immersed in a dark matter spike, Astrophys. J. 916, 116 (2021).
\bibitem{Arguelles} Arg$\ddot{\rm u}$elles C. R., Mestre M. F., Becerra-Vergara E. A., Crespi V., Krut A., Rueda  J. A., Ruffini  R., What does lie at the Milky Way centre? Insights from the S2 star orbit precession, Mon. Not. R. Astron. Soc. 511, L35 (2022).
\bibitem{Rubilar} Rubilar G. F., Eckart A., Periastron shifts of stellar orbits near the Galactic Center, Astron. Astrophys. 374, 95 (2001).
\bibitem{Nucita} Nucita, A. A., De Paolis F., Ingrosso G., Qadir A., Zakharov A. F., Sgr A*: A laboratory to measure the central black hole and stellar cluster parameters, Publ. Astron. Soc. Pac. 119, 349 (2007).
\bibitem{Zakharov} Zakharov A. F., Nucita A. A., de Paolis F., Ingrosso G., Apoastron shift constraints on dark matter distribution at the Galactic Center, Phys. Rev. D 76, 062001 (2007).
\bibitem{Dokuchaev} Dokuchaev V. I., Eroshenko Y. N., Physical laboratory at the center of the Galaxy, Physics-Uspekhi 58, 772 (2015).
\bibitem{Dokuchaev2} Dokuchaev V. I., Eroshenko Y. N., Weighing of the dark matter at the center of the galaxy, JETP Letters 101, 777 (2015).
\bibitem{Abuter2} Abuter R. {\it et al.} [GRAVITY Collaboration], The mass distribution in the Galactic Centre from interferometric astrometry of multiple stellar orbits, Astron. Astrophys. 657, L12 (2022).
\bibitem{Do} Do T. {\it et al.}, Relativistic redshift of the star S0-2 orbiting the Galactic center supermassive black hole, Science 365, 664 (2019).
\bibitem{Bahcall} Bahcall J., Wolf R. A., Star distribution around a massive black hole in a globular cluster, Astrophys. J. 209, 214 (1976).
\bibitem{Iorio} Iorio L., Long-term classical and general relativistic effects on the radial velocities of the stars orbiting Sgr A*, Mon. Not. R. Astron. Soc. 411, 453 (2011).
\bibitem{Generozov} Generozov A., Stone N. C., Metzger B. D., Ostriker J. P., An overabundance of black hole X-ray binaries in the Galactic Centre from tidal captures, Mon. Not. R. Astron. Soc. 478, 4030 (2018).
\bibitem{Plummer} Plummer H. C., On the problem of distribution in globular star clusters: (Plate 8.), Mon. Not. R. Astron. Soc. 71, 460 (1911).
\bibitem{Profumo} Profumo S., Giani L., Piattella O. F., An introduction to particle dark matter, Universe 5, 213 (2019).
\bibitem{Abuter1} Abuter R. {\it et al.} [GRAVITY Collaboration], Detection of the gravitational redshift in the orbit of the star S2 near the Galactic centre massive black hole, Astron. Astrophys. 615, L15 (2018).
\bibitem{Abuter3} Abuter R. {\it et al.} [GRAVITY Collaboration], A geometric distance measurement to the Galactic center black hole with 0.3\% uncertainty, Astron. Astrophys. 625, L10 (2019).
\bibitem{Steigman} Steigman G., Dasgupta B., Beacom J. F., Precise relic WIMP abundance and its impact on searches for dark matter annihilation, Phys. Rev. D 86, 023506 (2012).
\bibitem{Peibker2} Pei\ss ker F., Eckart A. \& Parsa M., S62 on a 9.9 yr orbit around Sgr A*, Astrophys. J. 889, 61, (2020).
\bibitem{Peibker3} Pei\ss ker F., Eckart A., Zaja\~cek M. Ali B. \& Parsa M., S62 and S4711: indications of a population of faint fast-moving stars inside the S2 orbit - S4711 on a 7.6 yr orbit around Sgr A*, Astrophys. J. 899, 50 (2020).
\bibitem{Peibker} Pei\ss ker F., Eckart A., Zaja\~cek M. \& Britzen S., Observation of S4716 -- a star with a 4 yr orbit around Sgr A*, Astrophys. J. 933, 49 (2022).
\end{thebibliography}

 
\begin{addendum}
\item The work described in this paper was partially supported by the Seed Funding Grant (RG 68/2020-2021R) and the Dean's Research Fund (activity code: 04628) from The Education University of Hong Kong.
\item[Data Availability Statement] The data that support the findings of this study are available from the corresponding author upon reasonable request.
\item[Competing Interests] The authors declare that he has no competing financial interests.
\item[Author Contributions] Analysis of the results and writing of the manuscript were performed by M. H. Chan. The numerical modelling was done by C. M. Lee. The checking of data and results was done by C. W. Yu.
\item[Correspondence] Correspondence and requests for materials should be addressed to Man Ho Chan.~(email: chanmh@eduhk.hk).
\end{addendum}
 

\end{document}